\documentclass[fleqn,usenatbib]{mnras}

\usepackage{newtxtext,newtxmath}

\usepackage[T1]{fontenc}
\usepackage{float}
\DeclareRobustCommand{\VAN}[3]{#2}
\let\VANthebibliography\thebibliography
\def\thebibliography{\DeclareRobustCommand{\VAN}[3]{##3}\VANthebibliography}
\newcommand\given[1][]{\:#1\vert\:}

\usepackage{graphicx}	
\usepackage{amsmath}	
\usepackage[skip=2pt]{caption}

\newcommand{\codefont}[1]{{\texttt{#1}}}

\title[The large-scale environment of SNeIa and CCSNe]{The large-scale environment of thermonuclear and core-collapse supernovae}

\author[Eleni Tsaprazi]{\parbox{16cm}{Eleni Tsaprazi,$^{1}$\thanks{E-mail: eleni.tsaprazi@fysik.su.se (Physics Dept., SU)}
Jens Jasche,$^{1}$
Ariel Goobar,$^{1}$
Hiranya V. Peiris,$^{1,2}$\\
Igor Andreoni,$^{3}$
Michael W. Coughlin,$^{4}$
Christoffer U. Fremling,$^{3}$
Matthew J. Graham,$^{3}$\\
Mansi Kasliwal,$^{3}$
Shri R. Kulkarni,$^{3}$
Ashish A. Mahabal,$^{3,5}$
Reed Riddle,$^{3}$
Jesper Sollerman$^{6}$\\
and Anastasios Tzanidakis$^{3}$}\\\\
$^{1}$The Oskar Klein Centre, Department of Physics, Stockholm University, Albanova University Center, SE-106 91 Stockholm, Sweden\\
$^{2}$Department of Physics \& Astronomy, University College London, Gower Street, London WC1E 6BT, UK\\
$^{3}$Division of Physics, Mathematics and Astronomy, California Institute of Technology, Pasadena, CA 91125, USA\\
$^{4}$School of Physics and Astronomy, University of Minnesota, Minneapolis, Minnesota 55455, USA\\
$^{5}$Center for Data Driven Discovery, California Institute of Technology, Pasadena, CA 91125, USA\\
$^{6}$The Oskar Klein Centre, Department of Astronomy, AlbaNova University Center, SE-106 91 Stockholm, Sweden\\
}

\date{Accepted XXX. Received YYY; in original form ZZZ}

\pubyear{2021}

\begin{document}

\label{firstpage}
\pagerange{\pageref{firstpage}--\pageref{lastpage}}
\maketitle

\begin{abstract}
The new generation of wide-field time-domain surveys has made it feasible to study the clustering of supernova (SN) host galaxies in the large-scale structure (LSS) for the first time. We investigate the LSS environment of SN populations, using 106 dark matter density realisations with a resolution of $\sim$ 3.8 Mpc, constrained by the 2M++ galaxy survey. We limit our analysis to redshift $z<0.036$, using samples of 498 thermonuclear and 782 core-collapse SNe from the Zwicky Transient Facility's {\sl Bright Transient Survey} and {\sl Census of the Local Universe} catalogues. We detect clustering of SNe with high significance; the observed clustering of the two SNe populations is consistent with each other. Further, the clustering of SN hosts is consistent with that of the Sloan Digital Sky Survey (SDSS) Baryon Oscillation Spectroscopic Survey (BOSS) DR12 spectroscopic galaxy sample in the same redshift range. Using a tidal shear classifier, we classify the LSS into voids, sheets, filaments and knots. We find that both SNe and SDSS galaxies are predominantly found in sheets and filaments. SNe are significantly under-represented in voids and over-represented in knots compared to the volume fraction in these structures. This work opens the potential for using forthcoming wide-field deep SN surveys as a complementary LSS probe.
\end{abstract}




\section{Introduction}

The large-scale structure (LSS) of the Universe forms a web-like pattern which consists of galaxies and intergalactic gas thought to trace a scaffolding of dark matter \citep{1978MNRAS.183..341W,1996Natur.380..603B}. Structures within this {\sl cosmic web} can be classified into voids, filaments, sheets and knots. Voids are regions with density lower than the cosmic mean, from which matter flows onto denser structures; sheets can be described as the boundaries of voids. Filaments are thread-like structures that intersect at knots. Classifications of the cosmic web can be based on different quantities such as density, peculiar velocity, tidal shear, resulting in differing sensitivity to the physical properties of cosmic structures depending on the classification scheme \citep[e.g.][]{2009MNRAS.396.1815F,2012MNRAS.425.2049H,2016MNRAS.458.1517F,2016JCAP...08..027L,2018MNRAS.473.1195L}.

Galaxies form in a complex interplay with their surroundings, and hence their properties are correlated with their cosmic web environments. Passive, old galaxies have been found to reside in dense large-scale structures, whereas young, star-forming galaxies have been observed in less dense structures \citep{1980ApJ...236..351D,Darvish_2014}.

Galaxies host supernova (SN) explosions which can be classified into two populations: thermonuclear explosions of white dwarfs in binary systems, also referred to as SNe Type Ia (SNeIa), and core-collapse SNe (CCSNe), which signal the demise of massive stars \citep{2000ARA&A..38..191H,2020arXiv200914157B}. The rates of SNe have been reported  to be dependent on the morphology of their host galaxy \citep[e.g.][]{2005A&A...433..807M}, its star-formation rate \citep[e.g.][]{2008ApJ...682L..25C,2012ApJ...755...61S} and stellar mass \citep[e.g.][]{2006ApJ...648..868S,2011MNRAS.412.1473L,2012ApJ...755...61S,2013MNRAS.430.1746G,2021MNRAS.506.3330W}. The use of CCSNe as tracers of the star formation history has been established with extensive observations carried out with the {\sl Hubble Space Telescope} \citep{2015ApJ...813...93S} and subsequent studies \citep[see][and references therein]{2020arXiv200805988S}.

In light of these correlations, a SN clustering signal in the LSS is expected, and the new generation of wide-field transient surveys opens up the possibility of measuring it for the first time. Previous studies have focused on the correlation of SN types with properties of SN host galaxies \citep[e.g.][]{2005A&A...433..807M,2008MNRAS.383.1121M,2005PASP..117..773V,2012ApJ...755...61S,2021MNRAS.506.3330W}, SN hosts with surrounding galaxies \citep{2008ApJ...682L..25C}, the cross-correlation of SNeIa with galaxy surveys \citep{2018arXiv180806615M} and in relation to galaxy density \citep{2009ApJ...704..687C}.

Upcoming SN samples can complement the study of galaxy clustering, as SNe can be found in host galaxies that are too faint to be resolved in galaxy surveys. In this work we investigate the connection of SNeIa and CCSNe with the LSS, as defined by the cosmic web structures which they trace. We use observations from the Zwicky Transient Facility \citep[ZTF;][]{2019PASP..131a8002B,2019PASP..131g8001G, 2020PASP..132c8001D,2019PASP..131a8003M} Bright Transient Survey \citep[BTS; ][]{2020ApJ...895...32F,2020ApJ...904...35P} and
Census of the Local Universe (CLU) catalogue \citep{2019ApJ...880....7C,2020arXiv200409029D} samples, in combination with large-scale density and peculiar velocity inferences using the Bayesian Origin Reconstruction from Galaxies (\codefont{BORG}) algorithm \citep{2019A&A...625A..64J}, constrained by the 2M$++$ galaxy survey data \citep{2011MNRAS.416.2840L}. We use the inferred velocity fields in combination with the measured redshifts of SNe to place them within the inferred density fields. We also classify the cosmic web structures within the inferred large-scale density fields using a tidal shear classifier \citep{2007MNRAS.375..489H}. Combining these processed data-products, we study the web-type distribution of SN host galaxies, and compare this with a representative set of galaxies in the same redshift range from the SDSS BOSS spectroscopic catalogue.

The paper is structured as follows: In Sec.~\ref{BORG_posteriors}, we describe the large-scale density and velocity fields used in this work. Sec.~\ref{tidal_shear} presents the cosmic web classification method. In Sec.~\ref{Objects}, we discuss the characteristics of the ZTF SN population studied in the present work and the SDSS BOSS DR12 galaxy sample that was used for comparison purposes. The statistical framework used to obtain the distribution of web-types for these different samples of sources is described in Sec.~\ref{Method}. Finally, in Secs.~\ref{Results} and \ref{Discussion} we present our results and conclusions, respectively.

\section{The large-scale density}\label{BORG_posteriors}

The \codefont{BORG} algorithm \citep{2013MNRAS.432..894J,2015JCAP...01..036J,2016MNRAS.455.3169L,2019A&A...625A..64J} provides inferences of the large-scale density and peculiar velocity field constrained by galaxy surveys. In the present study, we use the inference performed by \citet{2019A&A...625A..64J}. This inference was previously used by \citet{2018A&A...612A..31P} to study the large-scale environment of active galactic nuclei in a qualitatively similar manner to this work. The inferences are constrained using the 2M$++$ galaxy data \citep{2011MNRAS.416.2840L} within a cubic grid of side length 961.3 Mpc, a grid size of $256^3$ grid cells and resolution 3.8 Mpc. The observer is located at the centre of the inference domain. This field approach provides access to the structures along the line of sight (LOS) to the sources, which allows us to account for redshift uncertainties. Further, it contains all high-order statistics of the LSS and as a result, provides more information than cross-correlations between galaxies and SNe.

\codefont{BORG} fits a non-linear model of structure formation to galaxy survey data, exploiting high-order statistics of the LSS. For the 2M++ inference, the gravitational dynamics was implemented via a particle-mesh model \citep[][section 3.2]{2019A&A...625A..64J}. This forward-modelling approach allows \codefont{BORG} to infer the set of initial conditions consistent with the observed galaxy distribution, as well as the (non-linear) density and velocity fields corresponding to present-day structures.

\begin{figure}
\includegraphics[width=\columnwidth]{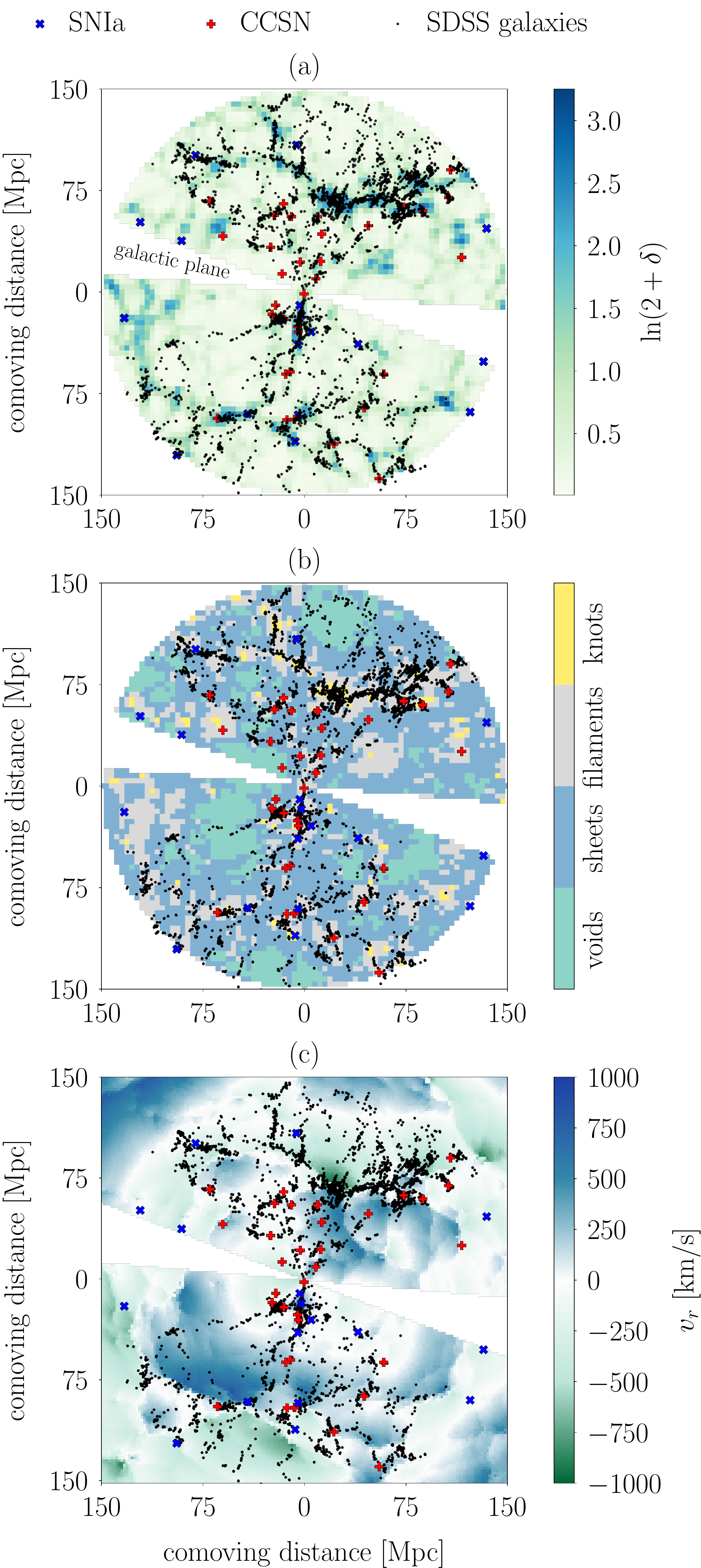}
\caption{(a) A density slice passing through the centre of the 3D density grid. The colour scale indicates the density contrast $\delta$, as $\mathrm{ln}(2+\delta)$, on a grid with a resolution of 3.8 Mpc. The axes represent the comoving distance from the observer (centre of the box). (b) The same slice for cosmic web structures. The colour scale indicates the four web types in the classification. (c) The same slice for the radial peculiar velocity passing through the observer in one realisation. The colour scale indicates the radial peculiar velocity in km\,s$^{-1}$ at 3.8 Mpc. The redshifts used for the overlay are the expectation values of the cosmological redshift posteriors of the sources. It can be seen that the web types in the middle panel are correlated with the density field and all populations trace the LSS.}  \label{fig:density_slice}
\centering
\end{figure}

The large-scale density and peculiar velocity fields are conditioned on the coordinates and magnitudes of the sources in the survey, the survey selection function, magnitude cuts and sky coverage. The resulting density and peculiar velocity posterior distributions are approximated by an ensemble of Markov Chain Monte Carlo samples. In the present study, we use 106 realisations of dark matter density and peculiar velocity fields. Our results are insensitive to the use of more realizations. The realisations are draws from the BORG posterior on initial conditions given by fitting a LCDM cosmology to the 2M++ data. Slices of the inferred dark matter density and radial peculiar velocity fields are shown in Fig.~\ref{fig:density_slice}a and Fig.~\ref{fig:density_slice}c, respectively. In particular, we use the peculiar velocity fields by \citet{2017JCAP...06..049L,2021A&A...646A..65M}, which were obtained on a grid of resolution $512^3$. These were constructed by re-binning the dark matter particle velocities in the original inference at $512^3$. In Fig.~\ref{fig:density_slice}b, the associated web types are indicated.

\section{Web type classification}\label{tidal_shear}

The identification of web types given the inferred large-scale dark matter density field is performed using the tidal shear classifier \citep{2007MNRAS.375..489H}. Here, the three eigenvalues $\lambda$ of the 3D tidal shear tensor of the gravitational potential $\Phi$ of the large-scale density field determine the web types on the $256^3$ grid. The shear-stress tensor $T_{ij}$ is a symmetric tensor defined as
\begin{equation}
T_{ij} = \frac{\partial^2\Phi}{\partial x_i \partial x_j},
\label{eq:Hessian}
\end{equation}
where $x_i$, ($\{i,j\} = \{1,2,3\}$) are comoving Cartesian coordinates. The resulting web types are then determined according to which eigenvalues $\lambda$ are greater than a threshold $\lambda_{\mathrm{th}}$, where their ordering is determined by the length of the corresponding eigenvectors. All three eigenvalues $< \lambda_{\mathrm{th}}$ result in a void, the two eigenvalues $\lambda_1, \lambda_2 < \lambda_{\mathrm{th}}$ result in a sheet, the eigenvalue $\lambda_1< \lambda_{\mathrm{th}}$ results in a filament, whereas no eigenvalue  $< \lambda_{\mathrm{th}}$ results in a knot. The threshold is indicative of when gravitational dynamics leads to the collapse of matter in the direction of the eigenvectors. A lower threshold leads to the classification of more gravitationally-collapsed structures. In the present study, the choice of threshold, $\lambda_{\mathrm{th}} = 0$, is determined by Lagrangian perturbation theory \citep{1996Obs...116...25C,2014MNRAS.437.3442H}. The resulting classification criteria are shown in Table~\ref{class_criterion}.

\begin{table}
\centering
\caption{The classification criteria for the tidal shear classifier. The value of the threshold $\lambda_{\mathrm{th}} = 0$ determines the web types traced. The choice of the classification threshold and the physical meaning of the structures is discussed in Sec.~\ref{tidal_shear}.}
\begin{tabular}{||c c ||}
 \hline
 Web structure & Eigenvalues \\ [0.5ex]
 \hline\hline
 voids & $\lambda_1$, $\lambda_2$, $\lambda_3$ < 0  \\
 sheets &$ \lambda_1$, $\lambda_2$ < 0,  $\lambda_3$ > 0 \\
 filaments & $\lambda_1$ <0, $\lambda_2$, $\lambda_3$ > 0  \\
 knots & $\lambda_1$, $\lambda_2$, $\lambda_3 $> 0 \\
 \hline
\end{tabular}
\label{class_criterion}
\vspace{-3mm}
\end{table}

The correspondence between the density field and the structures recovered by a tidal shear classifier depends on the value of the threshold $\lambda_{\mathrm{th}}$ \citep{2009MNRAS.396.1815F}. In contrast to density-based classifications, a tidal shear classifier provides a kinematic description of matter at any point. The resulting cosmic web types correlate with the density field. This can be better seen upon comparison of Fig.~\ref{fig:density_slice}a and Fig.~\ref{fig:density_slice}b. As is further shown in Table~\ref{tab:density}, voids and sheets are on average underdense, whereas filaments and knots are on average overdense. This classification is based on the density contrast, $\delta = \rho/\bar{\rho}-1$, where $\rho$ indicates density and $\bar{\rho}$ the average density in the observed volume. The mean density contrast which corresponds to a web type is the density contrast averaged over the grid cells of that web type and the realizations. Given the higher resolution of the peculiar velocity fields, we increase the resolution of the tidal tensor grid by upsampling by a factor of eight, to match that of the velocity fields.

\begin{table}
\centering
\caption{The mean density contrast of the four cosmic web types: voids, sheets, filaments and knots. Voids and sheets are underdense, whereas filaments and knots are overdense. }
\begin{tabular}{||c c ||}
 \hline
 Web structure & Mean density contrast \\ [0.5ex]
 \hline\hline
 voids & $-0.85$ \\
 sheets & $-0.56$ \\
 filaments & $1.03$  \\
 knots & $8.30$ \\
 \hline
\end{tabular}
\label{tab:density}
\vspace{-3mm}
\end{table}

\section{Supernova \& galaxy populations}\label{Objects}

\subsection{The ZTF SN sample}

\begin{figure}
\includegraphics[width=8.25cm]{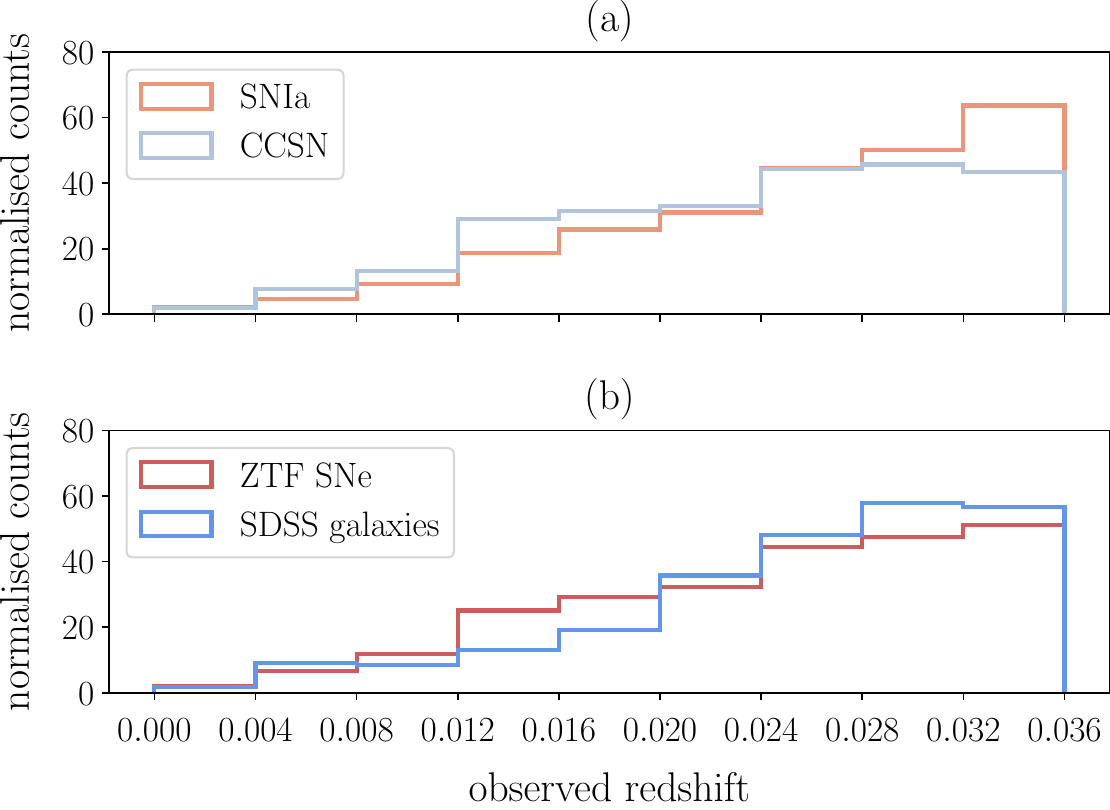}
\caption{Normalised distribution of the observed redshifts of (a) the SNIa and CCSN samples and (b) the entire ZTF sample and the SDSS BOSS DR12 sample. The similarity between the redshift distributions of the combined SNeIa and CCSNe sample, dubbed ZTF SNe, and the SDSS BOSS DR12 galaxies -- within Poisson uncertainty -- allows us to compare them directly without needing to account for redshift selection effects. The normalisation factor is the same for all distributions.} \label{fig:ZTF_SDSS_z_dist}
\end{figure}

\begin{figure}
\includegraphics[width=8.25cm]{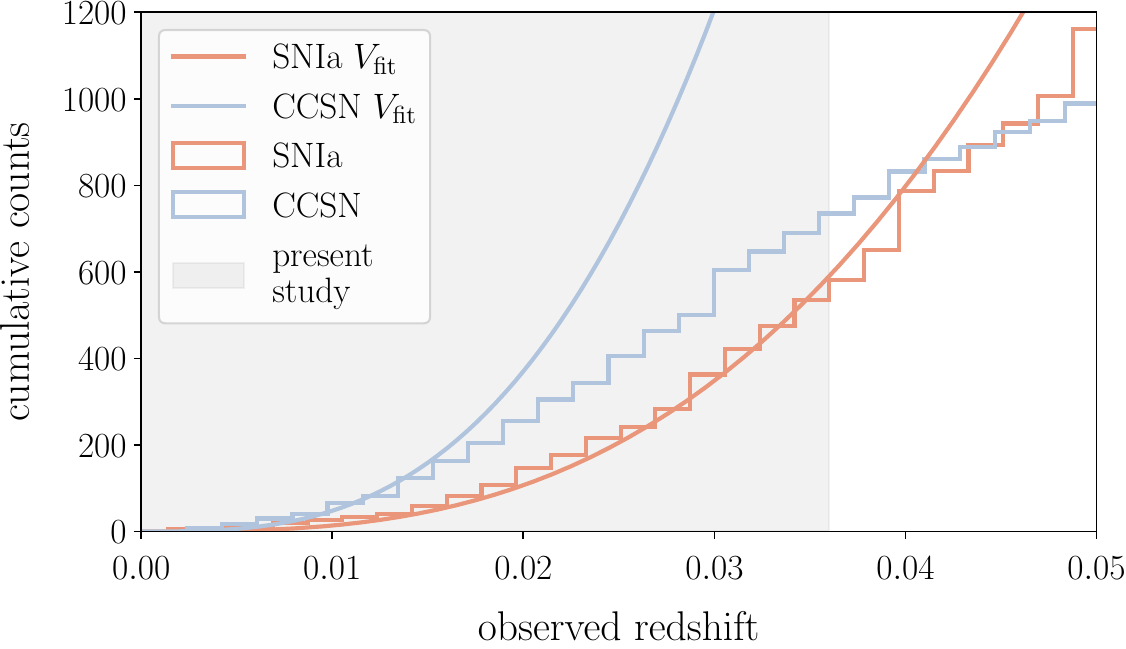}
\caption{Cumulative observed redshift distributions of the SNIa (orange) and CCSN (blue) samples up to redshift 0.05. The bin widths are the same for both samples. The curves represent a least-squares volume fit to the first 6 bins (CCSNe) and 20 bins (SNeIa) of the redshift histograms and indicate that the SNIa sample is complete up to the redshift cut of 0.036 used in the present study (grey shaded region).}  \label{fig:Ia_CC_completeness}
\end{figure}

The sample consists of spectroscopically-confirmed SNe, in particular 498 SNeIa and 782 CCSNe from Data Release 5 \citep{2019PASP..131a8002B,2020ApJ...895...32F, 2020ApJ...904...35P} of the ZTF BTS sample and events detected subsequently. The collection of the sample was performed using a BTS explorer query (\url{https://sites.astro.caltech.edu/ztf/bts/explorer.php}) on May 25$^\mathrm{th}$, 2021, without applying any cut apart from redshift. Our sample also contains 33 SNeIa and 153 CCSNe from the CLU experiment. There is no overlap between the BTS and CLU sample used in the present study. The observed redshift distributions of the two populations are shown in Fig.~\ref{fig:ZTF_SDSS_z_dist}a, after applying a redshift cut $z<0.036$, at the completeness limit of the SNIa population. The ZTF BTS does not provide an estimate of the uncertainty on the reported redshifts, but \citet{2020ApJ...895...32F} have reported very accurate host galaxy redshifts. In particular, as can be seen from \citet[fig. 5]{2020ApJ...895...32F}, the typical uncertainty on the difference between the redshift of the host galaxy and the SN is on the order of $\sim 0.005$. Therefore, we assume that three reported decimals on the SN redshifts indicate an uncertainty less than $\sigma_z$ = 0.005, which corresponds to a radial uncertainty of $\simeq 21$ Mpc. Under this assumption, the mean host redshift uncertainty for both SN populations is $\sim 10^{-4}$. Sources with three or more decimals are included in our sample, whereas sources with fewer decimals are discarded.
However, we do not discard SNe with no spectroscopic redshifts for their host galaxies. The SNe in our sample are spectroscopically identified and their redshifts are measured from the SN features. However, due to Doppler broadening, SN redshifts are not as accurate as host galaxy spectroscopic redshifts. Therefore, we use the latter when available. The redshift uncertainty is the same order of magnitude with the uncertainty due to the virial motion of the SN hosts within their haloes.

The completeness of the two samples is presented in Fig.~\ref{fig:Ia_CC_completeness} and indicates that the SNIa sample is complete in our selected redshift range. We estimate the redshift completeness using the least-squares fit function, $V_{\mathrm{fit}}$, which includes the effects of the Hubble expansion and the time dilation of the SN rate:

\begin{equation}
V_{\mathrm{fit}}(z) \propto \frac{1}{1+z}\int_0^z \frac{d_{\mathrm{c}}^2(z')}{(1+z')^3}dz',
\label{eq:completeness}
\end{equation}

\noindent where $d_{\mathrm{c}}$ is the comoving distance and $z$ the redshift. We consider the first 6 and 20 redshift bins for the CCSN and SNIa samples, respectively. The ratio of rates between the CCSN and SNIa populations, i.e. the ratio of the volume-fit slopes in Fig.~\ref{fig:Ia_CC_completeness}, is $\simeq$ 3.5. It is close to the rate 4.3 reported by \citet[fig. 9]{2020ApJ...904...35P} for the faint magnitude extrapolation, as the rate between the SNIa and CCSN events. Using Eq. \ref{eq:completeness}, the completeness ratio can be used as an indicator of the redshift at which the SN samples become incomplete.

\subsection{The SDSS BOSS DR12 galaxy sample}

We use $62$,$915$ spectroscopic galaxies from the SDSS BOSS DR12 catalogue \citep{2006AJ....131.2332G,2011AJ....142...72E,2013AJ....145...10D,2013AJ....146...32S,2015ApJS..219...12A} as a reference sample for comparison with SN host galaxies. Galaxies trace the large-scale density up to a bias \citep{1984ApJ...284L...9K}; this correlation can be seen by visual inspection of Fig.\ref{fig:density_slice}a. The cosmic web environments of this reference sample are compared to those of the SN sample, in order to investigate potential differences in the environments of galaxies that host SNe and those whose status as SN hosts is unknown. We choose the SDSS BOSS DR12 galaxy sample, as it has a similar redshift distribution to the ZTF sample, yielding results that can be compared to those of the SN populations directly, without introducing redshift selection effects. We select a subsample with $z<0.036$ from the SDSS BOSS DR12 catalogue to match the definition of the SN sample. The redshift distribution of the galaxies is compared with that of the SNe in Fig.~\ref{fig:ZTF_SDSS_z_dist}b.

\section{Method}\label{Method}

\begin{figure*}
\includegraphics[width=\textwidth]{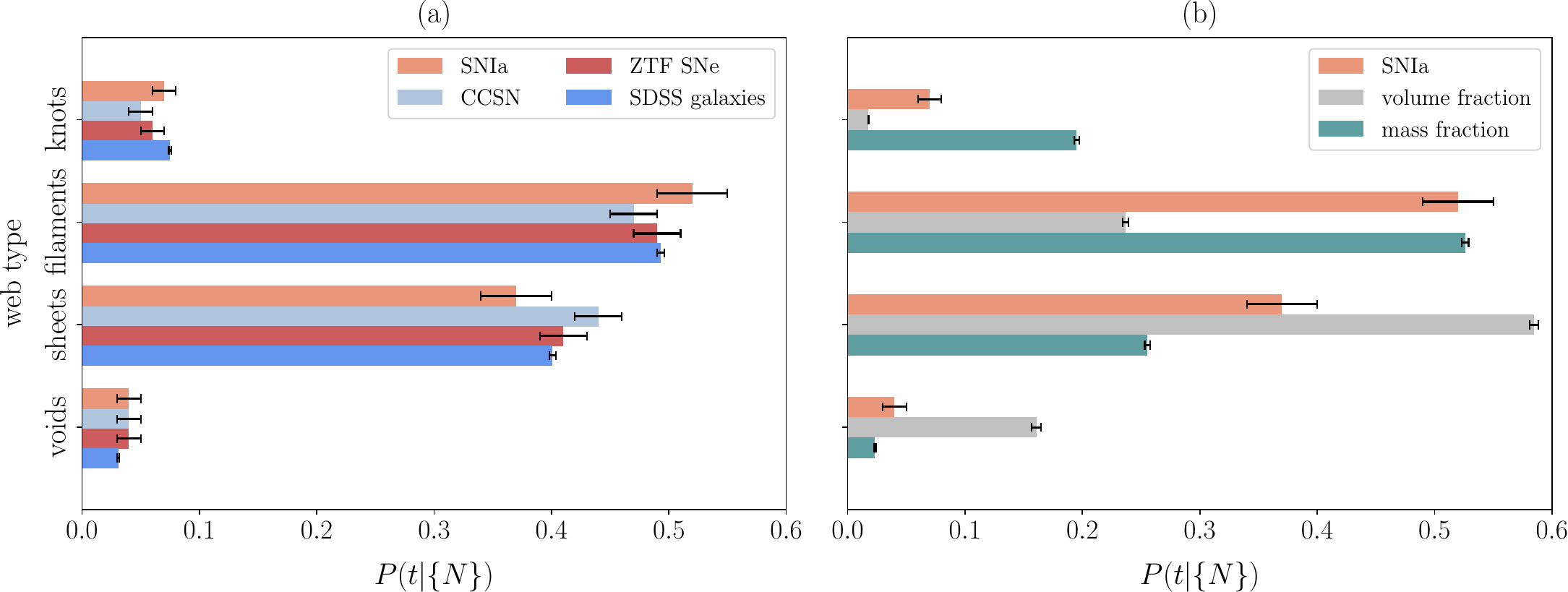}
\caption{(a) Web type posterior distributions for the SNIa, CCSN, ZTF and SDSS samples with 1$\sigma$ error bars. ZTF SNe represent the combined SNIa and CCSN samples. We find that SNeIa and CCSNe trace the LSS similarly, and that SNe trace the LSS like galaxies. Most sources trace sheets and filaments and a small fraction traces voids and knots. (b) Web type posterior distribution for the SNIa sample, compared to the web types in the observed volume and the mass distribution per web type. It can be seen that SNeIa are clustered and do not trace the mass and density in the observed volume.
}
\label{fig:Ia_vs_CC}
\end{figure*}

\subsection{Redshift-space distortions}

In order to associate the SNe with the cosmic web structures, it is necessary to transform the observed equatorial coordinates to comoving Cartesian coordinates in the inference grid. Redshift-space distortions contribute to a displacement between the observed and cosmological redshift \citep{1987MNRAS.227....1K}, to first order, as follows \citep[e.g.][]{2011ApJ...741...67D}:

\begin{equation}
	\label{eq:RSD}
	\hat{z}_n = z + \frac{\textbf{v}_\mathrm{h}\hat{\mathbfit{n}}}{c},
\end{equation}

\noindent where $\hat{z}_n$ is the observed redshift of a source $n$, $z$ the cosmological redshift, $\textbf{v}_\mathrm{h}$ the 3D peculiar velocity of the halo at the location of the source, $\hat{\mathbfit{n}}$ the unit LOS vector to the source and $c$ the speed of light. The observed redshifts are transformed from the heliocentric to the cosmic microwave background (CMB) frame using the corresponding transformation by \citet{2008ApJ...676..184T}, as the density and peculiar velocity fields are also in the CMB frame \citep{2019A&A...625A..64J}. The inferred peculiar velocity fields we use do not have the resolution to determine the virial velocity of the galaxies within their own halos, $\textbf{v}_\mathrm{vir}$. Hence we assume that $\textbf{v}_\mathrm{vir}$ contributes to the velocity dispersion per source $n$, as follows \citep{2001MNRAS.322..901S}:

\begin{equation}
    \sigma_{\mathrm{vir}_n} = 476  g_v [\Delta_\mathrm{nl}(z)E^2(z)]^{1/6}\left(\frac{M_{h_n}}{10^{15} M_\odot / h}\right)^{1/3} \mathrm{km\,s}^{-1},
    \label{virial}
\end{equation}
where $M_{h_n}$ is the halo mass of the source $n$, $g_v = 0.9$ is the growth rate,  $\Delta_\mathrm{nl}(z)=18\pi^2 + 60 x - 32 x^2$, with $x=\Omega_m(1+
z)^3/E^2(z) - 1$ and $ E(z) = \sqrt{\Omega_m(1 + z)^3 + (1 - \Omega_m)}$ \citep{2021A&A...646A..65M}. Throughout this study, we assume $H_0 = 70.5$ km$\mathrm{s}^{-1}\mathrm{Mpc}^{-1}$, $\Omega_m = 0.307$, $\Omega_b = 0.04825$, $\Omega_\Lambda = 0.693$ and $w = -1$, consistent with \citet{2019A&A...625A..64J}. The halo masses of the SDSS galaxies are derived from the corresponding stellar masses in the SDSS catalogue via \citet[eq.~6]{2020A&A...634A.135G} for redshift $z = 0$. The stellar masses of the ZTF SNe are derived via \citet[eq.~3]{2021arXiv210506236J} and then transformed to halo masses. In the cases where host photometry is not available or the stellar masses fall outside the range reported in \citet{2020A&A...634A.135G}, which constitute 14\% of the SNeIa and 28\% of the CCSNe, we draw random stellar masses from the distribution of stellar masses of ZTF hosts in \citet{Dhawan_2021}. Discarding these sources does not change our results significantly.

\subsection{The cosmic web type posterior}

We infer the cosmic web structures traced by SNeIa, CCSNe and galaxies whose status as SN hosts is unknown. In formulating the web type posterior, we assume a uniform cosmological redshift prior and ignore correlations between neighbouring peculiar velocity grid cells. Furthermore, we assume that all  realisations contribute equally to the web type posterior, and further, that the uncertainty on the angular position of the sources in the sky is negligible. We also assume that the cosmic web types are independent of the peculiar velocities, since the former were derived from the density field.

We assume a Gaussian likelihood whose mean is given by Eq. \ref{eq:RSD}. The total variance consists of the redshift uncertainty, $\sigma^2_z$, and the velocity dispersion due to the virial motions of galaxies within their halos. Under these assumptions, the posterior for the web types $t$, $t\in\{1,...,4\}$ given a source population $\{N\}$ can be written as a marginal over the LOS grid cells, the realisations and the sources:
\begin{eqnarray}
    \mathcal{P}(t|\{N\})  &\propto& \sum_{n=1}^N L_{n,k} \sum_{k=1}^K \sum_{j=1}^J \mathcal{N}\left(\hat{z}_n\given[\Big] z_{k}+\frac{v_{k,j}}{c},\sigma^2_n\right)\\ \nonumber
    &\times& \delta^\mathrm{D}(t-t_{k,j}),
    \label{posterior}
\end{eqnarray}
where $\delta^D$ is the Dirac delta function and $\mathcal{N}(x|\mu,\sigma)$ is a Gaussian distribution with mean $\mu$ and standard deviation $\sigma$. The index $j\in\{1,...,J\}$ denotes the  realisations and the index $n\in\{1,...,N\}$ denotes the sources in the population. The indices $k\in\{1,...,K\}$ denote the grid cells in the observed volume. The $L_{n,k}$ operator is 1 if a grid cell lies along the LOS to a source and 0 otherwise.

We take the total uncertainty, $\sigma$, to consist of two independent components: the uncertainty of the observed redshift estimate and the virial motion of the galaxy within its halo. The two are combined as
\begin{equation}
    \sigma^2_n = \sigma^2_{z, n} + \left(\frac{\sigma_{\mathrm{vir}_n}}{c}\right)^2 .
\end{equation}
Finally, we assume a $\sqrt{N}$ Poisson uncertainty on the web-type counts given a source population. This uncertainty model was verified using jackknifing.

To validate our algorithm, we infer the radial peculiar velocity posterior for NGC 4993, which has been studied by \citet{2021A&A...646A..65M} and references therein. When we consider only the peculiar motion of the halo, we find a posterior mean of $\bar{v}_p = 354 \pm 75$ km\,s$^{-1}$. When we add the peculiar velocity uncertainty due to the virial motion of the galaxy within its halo, we recover $\bar{v}_p = 357 \pm 84$ km\,s$^{-1}$. In doing so, we use Eq.~\ref{virial} for a halo mass of $10^{12} M_\odot$ \citep{2017ApJ...848L..30P,2020A&A...634A..73E}. Our results agree with \citet{2021A&A...646A..65M} both with and without the virial velocity component. The velocity dispersion in our approach is slightly smaller, as we have not considered correlations between neighbouring peculiar velocity grid cells.

\section{Results}\label{Results}

We now present the cosmic web type posterior distributions for SNeIa and CCSNe compared with the SDSS galaxy sample. We also consider two reference fractions: the web type fraction, and the mass fraction per web type, both in the entire observed volume. We will refer to these distributions as the volume and mass fractions, respectively. The SDSS galaxy sample is used to investigate potential differences in the cosmic environments traced by galaxy and SN surveys. The volume fraction is used to show that SNeIa are clustered. The mass fraction is used to probe how SNeIa trace the total mass in the observed volume.

A summary of our results is given in Table~\ref{tab:results} and illustrated in Fig.~\ref{fig:Ia_vs_CC}a. We find similar web type distributions for SNeIa and CCSNe, which trace mostly underdense sheets and overdense filaments. We further find that the combined sample of SNeIa and CCSNe traces the LSS similarly to the SDSS galaxy sample in the same redshift range. These results demonstrate that SNe, like galaxies, are clustered, as their distribution across the web types is consistent with that of the SDSS galaxies, which are biased tracers of the dark matter density.

Finally, we compare the volume-limited SNIa\footnote{For CCSNe, any comparison to the volume and mass fractions should take into account the fact that the sample is not volume-limited.} sample to the volume and mass fractions. The results are shown in Table~\ref{tab:results} and illustrated in Fig.~\ref{fig:Ia_vs_CC}b. The SNIa web-type distribution is significantly different from the fractional volume occupied by the different web types. This confirms that SNeIa are strongly clustered; The comparison with the fractional mass suggests that SNeIa are biased with respect to the background mass distribution. The fraction of SNeIa does not match the mass distribution in knots and sheets. This may be due to a combination of different stellar age distributions and mass-to-light ratios, with knots being richer in dark matter compared to the luminous matter tracing stellar density \citep{2021MNRAS.506.3330W}.

\begin{table}
\caption{The web type fraction for the SNeIa, CCSNe and SDSS galaxies and the fractional volume and mass per web type. ZTF SNe represent the combined SNIa and CCSN samples. We find at high significance that SNe are clustered.}
\begin{tabular}{|p{1.65cm}|p{1.2cm}|p{1.2cm}|p{1.2cm}|p{1.2cm}|}
\hline
Population    & voids         & sheets        & filaments     & knots       \\[0.5ex]
\hline \hline
SNeIa          & 0.04±0.01   & 0.37±0.03   & 0.52±0.03   & 0.07±0.01 \\ \hline
CCSNe        & 0.04±0.01   & 0.44±0.02   & 0.47±0.02   & 0.05±0.01 \\ \hline
ZTF SNe       & 0.04±0.01   & 0.41±0.02   & 0.49±0.02   & 0.06±0.01 \\ \hline
SDSS galaxies & 0.031±0.001 & 0.401±0.003 & 0.493±0.003 & 0.075±0.001 \\ \hline
Volume & 0.161±0.004 & 0.584±0.004 & 0.237±0.002 & 0.018±0.001
\\ \hline
Mass & 0.023±0.001 & 0.255±0.003 & 0.526±0.003 & 0.196±0.002
\\ \hline
\end{tabular}
\label{tab:results}
\end{table}

\section{Discussion \& Conclusions}\label{Discussion}

We have performed a study of the large-scale environments of SNIa and CCSN populations in the low redshift Universe, and compared the results with a reference galaxy sample at $0<z<0.036$. We found with high significance that SNeIa and CCSNe are clustered, with the level of clustering being similar for both SN types. We found an $\sim11\sigma$ detection of clustering of SNeIa in voids, $\sim7\sigma$ in sheets, $\sim9\sigma$ in filaments and $\sim5\sigma$ in knots, by comparing the SNIa web-type fractions with the corresponding volume fractions. We showed that the ZTF SN survey traces the same LSS structures as galaxy surveys. We further found that SNe and galaxies, when used as tracers of the LSS, primarily trace sheets and filaments. We also showed that SNeIa are biased tracers of the mass distribution in the observed volume. In particular, we find that the fraction of SNIa is lower (higher) than the mass fraction in knots (sheets), an effect which we attribute to a density-dependence of the baryonic to dark matter ratio.

Our finding that SNe are highly clustered potentially has major implications for the clustering assumptions in previous SN survey simulations, where the SN distribution is typically assumed to be uniform \citep{2002A&A...392..757G,2019JCAP...10..005F}. More broadly, our results indicate the potential for using SNe as complementary LSS probes of cosmology. Structure growth analyses, so far exploiting SNe as standard candles \citep[e.g.][]{2020PhRvD.101b3516K,2020arXiv200109095G}, can further consider SNe as tracers of the large-scale density at redshifts where galaxies are too faint to be resolved \citep[e.g.][eq. 1]{2020MNRAS.498.2703B}. Our framework includes the study of SNeIa from ZTF and the LSS inferences from \codefont{BORG}, naturally incorporating non-linear effects. Hence, it has the potential to provide important tests of gravity once the ZTF Hubble diagram becomes available.

The next-generation galaxy surveys, while reaching much deeper magnitude limits, will still predominantly resolve the brighter galaxy populations. SNe are bright point sources and each SN subtype spans a narrower luminosity range than galaxies. Therefore, SNe can be used to alleviate potential redshift-dependent biases in the probed galaxy demographics. Coming surveys will further provide significantly larger SN samples. The extended ZTF survey is expected to detect $\sim 10,000$ low-redshift SNeIa and a comparable number of CCSNe \citep{2019PASP..131g8001G,2020ApJ...904...35P}. The {\sl Vera C. Rubin Observatory}'s {\sl Legacy Survey of Space and Time} is expected to detect $\sim 50,000$ SNeIa per year \citep{2009arXiv0912.0201L}. The {\sl Nancy Grace Roman Space Telescope} will deliver $\sim 1,000$ SNe at high redshifts \citep{2021arXiv211103081R}. Such sample sizes combined with complementary galaxy samples, as biased tracers of the LSS, could provide constraints on the growth of structure over cosmic time and gravity \citep{2017ApJ...847..128H,2019BAAS...51c.140K}.

In order to harness the full power of such large SN catalogues, improvements must be made in modelling the LSS within \codefont{BORG}, to construct deep, higher-resolution LSS inferences. The latter will enable a refined association of the sources with their environmental properties. Such an extension of our analysis to smaller scales can be exploited for astrophysical and cosmological studies \citep[e.g.][]{2012IAUS..279..183A,2012A&A...545A..96M,2013A&A...560A..66R,2015PASA...32...19A,2018ApJ...854...24K,2018A&A...615A..68R} with upcoming time-domain surveys.

\section*{Author contributions}

The main roles of the authors were, using the CRediT (Contribution Roles Taxonomy) system (\url{https://authorservices.wiley.com/author-resources/Journal-Authors/open-access/credit.html}):

\textbf{E.T.:} conceptualisation, methodology, software, formal analysis, writing - original draft.

\textbf{J.J.:} data, software, resources, validation, supervision, writing - review, funding acquisition.

\textbf{A.G.:} data, conceptualisation, writing - review \& editing, supervision, funding acquisition.

\textbf{H.V.P.:} conceptualisation, methodology, software, validation, writing - review \& editing, funding acquisition.

In alphabetical order:

\textbf{I.A.}, \textbf{M.W.C.}, \textbf{C.U.F.}, \textbf{M.J.G.}, \textbf{M.K.}, \textbf{S.R.K.}, \textbf{A.A.M.}, \textbf{R.R.}, \textbf{J.S.}, \textbf{A.T.:} data curation

\section*{Acknowledgements}

We thank the anonymous referee for the useful comments. ET thanks Guilhem Lavaux and Suvodip Mukherjee for helpful discussions and access to the peculiar velocity data; and Florent Leclercq and Natalia Porqueres for providing test data during the early stages of this work. HVP thanks Daniel Mortlock for helpful discussions regarding the statistical formalism. This work has been enabled by support from the research project grant ‘Understanding the Dynamic Universe’ funded by the Knut and Alice Wallenberg Foundation under Dnr KAW 2018.0067. JJ acknowledges support by the Swedish Research Council (VR) under the project 2020-05143 -- "Deciphering the Dynamics of Cosmic Structure". AG acknowledges support from the Swedish Research Council under Dnr VR 2020-03444. HVP was partially supported by the research project grant ``Fundamental Physics from Cosmological Surveys'' funded by VR under Dnr 2017-04212. This work uses products of the Aquila Consortium (\url{https://aquila-consortium.org}).

This study is based on observations obtained with the Samuel Oschin 48-inch Telescope at the Palomar Observatory as part of the Zwicky Transient Facility project. ZTF is supported by the National Science Foundation under Grant No. AST-1440341 and a collaboration including Caltech, IPAC, the Weizmann Institute for Science, the Oskar Klein Centre at Stockholm University, the University of Maryland, the University of Washington, Deutsches Elektronen-Synchrotron and Humboldt University, Los Alamos National Laboratories, the TANGO Consortium of Taiwan, the University of Wisconsin at Milwaukee, and Lawrence Berkeley National Laboratories. Operations are conducted by COO, IPAC, and UW.

This work makes use of public data from the Sloan Digital Sky Survey (SDSS). Funding for SDSS-III has been provided by the Alfred P. Sloan Foundation, the Participating Institutions, the National Science Foundation, and the U.S. Department of Energy Office of Science. The SDSS-III web site is \url{http://www.sdss3.org/}.

SDSS-III is managed by the Astrophysical Research Consortium for the Participating Institutions of the SDSS-III Collaboration including the University of Arizona, the Brazilian Participation Group, Brookhaven National Laboratory, Carnegie Mellon University, University of Florida, the French Participation Group, the German Participation Group, Harvard University, the Instituto de Astrofisica de Canarias, the Michigan State/Notre Dame/JINA Participation Group, Johns Hopkins University, Lawrence Berkeley National Laboratory, Max Planck Institute for Astrophysics, Max Planck Institute for Extraterrestrial Physics, New Mexico State University, New York University, Ohio State University, Pennsylvania State University, University of Portsmouth, Princeton University, the Spanish Participation Group, University of Tokyo, University of Utah, Vanderbilt University, University of Virginia, University of Washington, and Yale University.

\section*{Data availability}

The SN data underlying this article are available in \url{https://nextcloud.fysik.su.se/s/nnwsFaGFeKqpx7Q}.

\bibliographystyle{mnras}
\bibliography{main}

\bsp
\label{lastpage}
\end{document}